\documentclass[pdflatex,sn-nature]{sn-jnl}%

\usepackage{graphicx}%
\usepackage{multirow}%
\usepackage{amsmath,amssymb,amsfonts}%
\usepackage{amsthm}%
\usepackage{mathrsfs}%
\usepackage[title]{appendix}%
\usepackage{xcolor}%
\usepackage{textcomp}%
\usepackage{manyfoot}%
\usepackage{booktabs}%
\usepackage{longtable}%
\usepackage{algorithm}%
\usepackage{algorithmicx}%
\usepackage{algpseudocode}%
\usepackage{listings}%

\graphicspath{{figures/}}%

\usepackage[T1]{fontenc}%
\IfFileExists{newtxtext.sty}{%
  \usepackage{newtxtext}%
  \let\SNsaveopenbox\openbox%
  \let\openbox\relax%
  \usepackage{newtxmath}%
  \ifx\SNsaveopenbox\relax\else\let\openbox\SNsaveopenbox\fi%
}{%
  \usepackage{mathptmx}%
}%

\usepackage[mathlines,left]{lineno}%
\makeatletter
\newcommand*\SN@patchAmsMathEnv[1]{%
  \expandafter\let\csname SNold#1\expandafter\endcsname\csname #1\endcsname
  \expandafter\let\csname SNoldend#1\expandafter\endcsname\csname end#1\endcsname
  \renewenvironment{#1}%
    {\linenomath\csname SNold#1\endcsname}%
    {\csname SNoldend#1\endcsname\endlinenomath}}%
\newcommand*\SN@patchBothAmsMathEnvs[1]{%
  \SN@patchAmsMathEnv{#1}\SN@patchAmsMathEnv{#1*}}%
\AtBeginDocument{%
  \SN@patchBothAmsMathEnvs{equation}%
  \SN@patchBothAmsMathEnvs{align}%
  \SN@patchBothAmsMathEnvs{flalign}%
  \SN@patchBothAmsMathEnvs{alignat}%
  \SN@patchBothAmsMathEnvs{gather}%
  \SN@patchBothAmsMathEnvs{multline}%
}%
\makeatother

\usepackage{etoolbox}%
\makeatletter
\patchcmd{\@maketitle}
  {\removelastskip\vskip20pt\nointerlineskip}
  {\removelastskip\vskip4pt\nointerlineskip}
  {}{\PackageError{PhysECD}{Could not adjust the title spacing}{Check the title layout in sn-jnl.cls.}}%
\makeatother

\makeatletter
\renewcommand{\abstractfont}{%
  \reset@font\fontsize{9bp}{11bp}\selectfont
  \leftskip=20pt\rightskip=20pt\parfillskip=0pt plus 1fil}%
\makeatother

\usepackage{setspace}%

\theoremstyle{thmstyleone}%

\theoremstyle{thmstyletwo}%

\theoremstyle{thmstylethree}%

\newcommand{\muv}{\boldsymbol{\mu}}
\newcommand{\mv}{\mathbf{m}}
\newcommand{\rr}{\mathbf{r}}

\newcommand*{\includegraphicsTG}[2][\textwidth]{%
  \ifdefined\pdfximage
    \pdfximage attr{/Group <</S /Transparency /CS /DeviceRGB>>}{#2}%
    \setbox0=\hbox{\pdfrefximage\pdflastximage}%
    \resizebox{#1}{!}{\box0}%
  \else
    \includegraphics[width=#1]{#2}%
  \fi}%

\newcommand{\PCC}{\mathrm{PCC}}

\begin{document}

\title[PhysECD]{Parity-equivariant prediction of conformer-resolved electronic circular dichroism spectra accelerates absolute configuration assignment}

\author[1,2]{\fnm{Yi} \sur{Jiang}}
\equalcont{These authors contributed equally to this work.}

\author[1,3]{\fnm{Letian} \sur{Chen}}
\equalcont{These authors contributed equally to this work.}

\author[1]{\fnm{Runhan} \sur{Shi}}

\author[4,5,6,7]{\fnm{Lea} \sur{Bae}}

\author[8]{\fnm{Zheng} \sur{Ling}}

\makeatletter
\g@addto@macro\artauthors{\def\authorsep{{\au@and}\break}}
\makeatother

\author[4,5,6,7,9]{\fnm{Dong Ha} \sur{Kim}}

\author*[3,10]{\fnm{Tong} \sur{Zhu}}\email{tzhu@lps.ecnu.edu.cn}

\author*[1]{\fnm{Yang} \sur{Yang}}\email{alayman@sjtu.edu.cn}

\affil[1]{\orgdiv{Institute of Artificial General Intelligence},
\orgdiv{School of Computer Science}, \orgname{Shanghai Jiao Tong
University}, \orgaddress{\city{Shanghai}, \postcode{200240},
\country{China}}}

\affil[2]{\orgdiv{Global Institute of Future Technology}, \orgname{Shanghai
Jiao Tong University}, \orgaddress{\city{Shanghai}, \postcode{200240},
\country{China}}}

\affil[3]{\orgname{Shanghai Innovation Institute},
\orgaddress{\city{Shanghai}, \postcode{200231}, \country{China}}}

\affil[4]{\orgdiv{Department of Chemistry and Nanoscience, Division of Molecular Life \& Chemical Sciences, College of Natural Sciences},
\orgname{Ewha Womans University},
\orgaddress{\city{Seoul},
\postcode{03760}, \country{Korea}}}

\affil[5]{\orgdiv{Graduate Program in Innovative Biomaterials Convergence},
\orgname{Ewha Womans University},
\orgaddress{\city{Seoul},
\postcode{03760}, \country{Korea}}}

\affil[6]{\orgdiv{Basic Sciences Research Institute (Priority Research Institute)},
\orgname{Ewha Womans University},
\orgaddress{\city{Seoul},
\postcode{03760}, \country{Korea}}}

\affil[7]{\orgdiv{Institute for Multiscale Matter and Systems},
\orgname{Ewha Womans University},
\orgaddress{\city{Seoul},
\postcode{03760}, \country{Korea}}}

\affil[8]{\orgdiv{School of Chemistry and Chemical Engineering},
\orgname{Shanghai Jiao Tong University}, \orgaddress{\city{Shanghai},
\postcode{200240}, \country{China}}}

\affil[9]{\orgdiv{College of Medicine},
\orgname{Ewha Womans University},
\orgaddress{\city{Seoul},
\postcode{07804}, \country{Korea}}}

\affil[10]{\orgdiv{Shanghai Frontiers Science Center of Molecule Intelligent
Syntheses}, \orgdiv{School of Chemistry and Molecular Engineering},
\orgname{East China Normal University}, \orgaddress{\city{Shanghai},
\postcode{200062}, \country{China}}}

\abstract{Electronic circular dichroism (ECD) is a key experimental probe for assigning the absolute configuration of chiral molecules, but interpreting measured spectra often requires costly time-dependent density functional theory (TDDFT) calculations. The few existing machine learning methods for ECD prediction are faster, but their accuracy still falls well short of that required for experimental applications. Here we introduce PhysECD, a parity-aware E(3)-equivariant framework that builds the spectrum from its underlying quantum-chemical quantities, ensuring correct chiroptical symmetry by construction. To improve generalization and support experimental applications, we construct SCECD, a conformer-resolved dataset of 578,104 computed spectra spanning diverse natural products with one to five stereocentres. PhysECD predicts ECD spectra from conformer geometries in milliseconds, achieving state-of-the-art accuracy on the CMCDS benchmark and SCECD. Validation against 77 experimental spectra further demonstrates that PhysECD can serve as a fast alternative to TDDFT for assigning the absolute configuration of complex chiral molecules.}

\maketitle

\section{Introduction}\label{sec:intro}

Chirality is a fundamental structural property in organic molecules, pharmaceuticals and functional materials. 
Although enantiomers share the same molecular composition and generally indistinguishable physicochemical properties, they can exhibit profound differences in biological activity, pharmacological efficacy, metabolism and toxicity~\citep{chirality_1, noyori2002, chirality_2, brooks2011, smith2009}. 
Consequently, the rapid and reliable assignment of absolute configuration remains a critical issue in asymmetric synthesis, natural product elucidation, drug discovery and materials design~\citep{nicolaou2005, yoo2016}.

Electronic circular dichroism (ECD) measures the differential absorption of left- and right-circularly polarized light by chiral molecules and thus provides spectral information directly related to absolute configuration~\citep{berova2007, mandi2019}.
Because the spectra of two enantiomers are exact mirror images of each other, the technique is inherently sensitive to handedness.
In a typical structure elucidation, the relative configuration is first established by nuclear magnetic resonance (NMR) spectroscopy or X-ray crystallography, leaving two enantiomers to distinguish.
Theoretical ECD spectra are then obtained through conformational search, geometry optimization, time-dependent density functional theory (TDDFT) calculations, Boltzmann weighting and Gaussian broadening~\citep{pescitelli2016, crawford2007current, superchi2018}.
The absolute configuration is assigned to the enantiomer whose calculated spectrum best matches the measured one~\citep{specdis}.
However, TDDFT is computationally expensive. A single calculation can consume hours of processor time, and its cost scales steeply with system size~\citep{rungegross1984, gross2012introduction}.
It must also be repeated for every low-energy conformer, so one flexible natural product can require tens of processor-hours.
The cost increases further when the relative configuration cannot be fixed, for example when stereocentres are separated by flexible linkers or lack diagnostic nuclear Overhauser effect contacts~\citep{bifulco2007}.
In such cases, each candidate diastereomer requires its own conformer ensemble.
A molecule with five stereocentres, for example, can require up to 16 such ensembles.

Machine learning promises to remove this computational bottleneck, and it has already transformed the prediction of most molecular spectra.
Models trained on quantum-chemical datasets now reproduce infrared (IR), Raman, ultraviolet--visible (UV--vis) and NMR spectra accurately at a small fraction of the cost of theoretical calculations~\citep{gastegger2017, shiftml, ghosh2019, painn, westermayr2021, detanet, xu2025pretrained}. 
However, the prediction of ECD spectra has not yet achieved the same level of maturity.
Only a few machine learning models have been developed for this task.
ECDFormer~\citep{ecdformer} pioneered this direction by decoupling the spectrum into peak properties and predicting them with a transformer~\citep{vaswani2017}, achieving large speedups over TDDFT.
ChiDeK~\citep{chidek} encoded the configuration of each stereocentre through chiral determinant kernels, outperforming ECDFormer in the ECD prediction task.
Yet, the accuracy of existing models still falls well short of what experimental applications require.

\begin{figure}[t]
\centering
\includegraphicsTG[\textwidth]{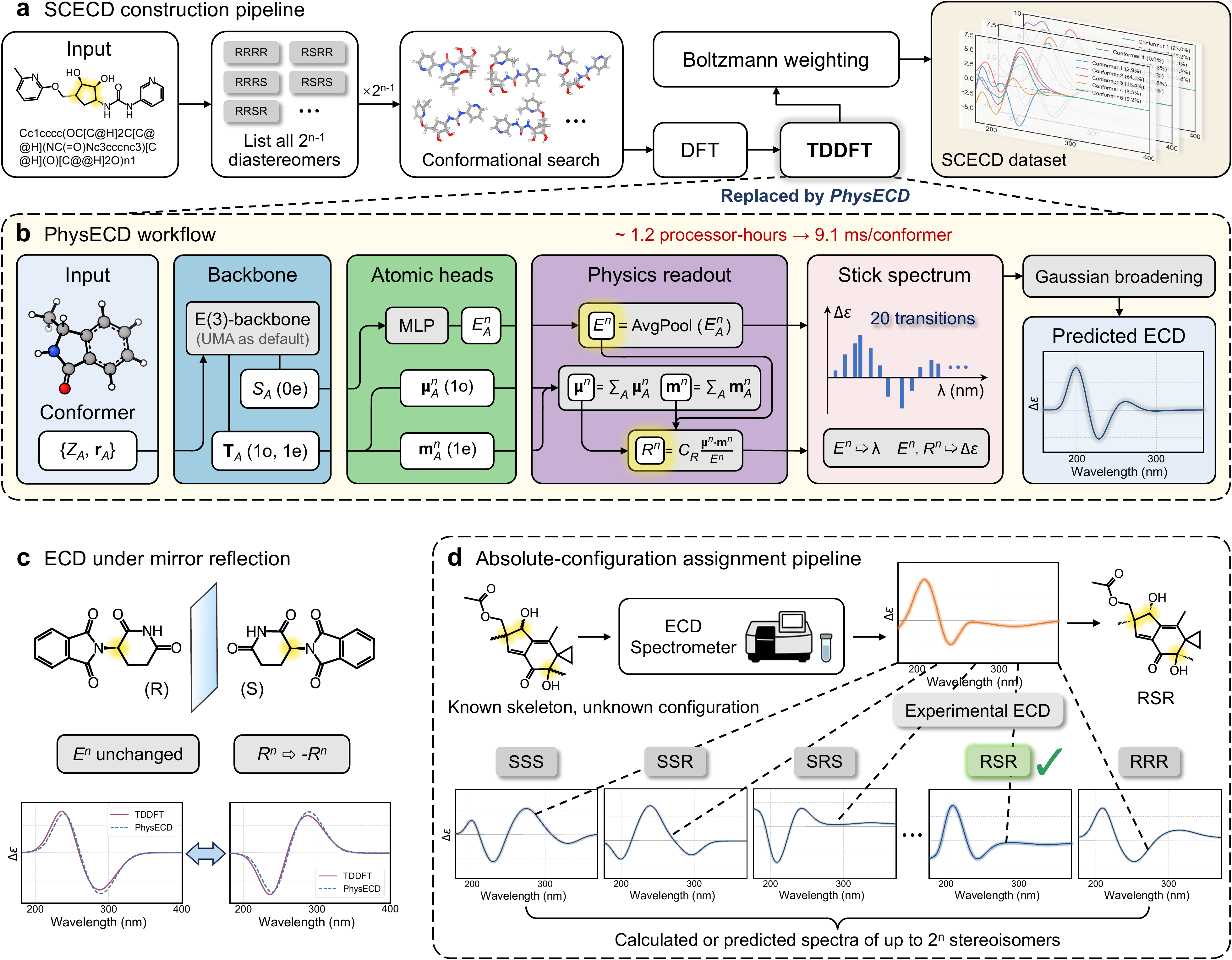}
\caption{\textbf{Overview of PhysECD.} \textbf{a}, Construction of the SCECD
dataset. For every parent molecule, the $2^{n-1}$ diastereomers of its $n$
stereocentres are enumerated. Each is carried through a conformational
search, a DFT geometry optimization and a TDDFT calculation of the 20 lowest
excited states. The conformer spectra and the electronic energies that set
their Boltzmann weights constitute the released dataset. \textbf{b}, The
PhysECD architecture, which replaces the TDDFT step. An E(3)-equivariant backbone maps the atomic numbers $Z_A$ and coordinates
$\rr_A$ to per-atom invariant scalars $S_A$ ($0e$) and equivariant
tensors $T_A$ ($1o$, $1e$). Atomic heads turn these into per-state excitation energies ($0e$) and the electric ($1o$) and magnetic ($1e$) transition dipoles. A parameter-free readout pools them into molecular-level quantities and generates the rotatory strengths. A closed-form Gaussian broadening converts the discrete transitions into a continuous spectrum. \textbf{c}, The ECD spectra of
a pair of enantiomers are mirror images of one another about the wavelength
axis. Reflection leaves every excitation energy unchanged and reverses the sign
of every rotatory strength. The two spectra therefore share one shape and differ only
in sign. \textbf{d}, Pipeline for assigning absolute configuration. The
measured spectrum is compared with the calculated or predicted spectra of candidate configurations, and the closest match identifies the configuration.}
\label{fig:arch}
\end{figure}

Two obstacles have hindered progress in this field, one in modelling and the other in data.
First, ECD exhibits a unique symmetry distinct from other spectroscopic techniques.
Enantiomers have identical IR, Raman, UV--vis and NMR spectra, whereas their ECD spectra are exactly inverted in sign.
A reliable predictor must reproduce this inversion, but existing models treat the spectrum as an abstract signal, unconstrained by the physics that produces it.
Nothing in these models ensures that enantiomers yield sign-inverted spectra.
Second, the available ECD datasets limit generalization and practical use in two ways.
They are dominated by molecules with a single stereocentre~\cite{cmcds}, yet the molecules whose configuration matters in practice usually carry several.
Of the 158 chiral small-molecule drugs approved by the US Food and Drug Administration over the past decade, 63\% carry more than one stereocentre and 38\% carry three or more~\citep{blakley2026}.
These datasets also represent each molecule by a single geometry, even though a measured spectrum is a Boltzmann average over multiple conformers.
To our knowledge, no existing public ECD dataset covers both multiple stereocentres and conformer ensembles.

Here, we address both obstacles. On the modelling side, we introduce \textbf{PhysECD}, a parity-aware E(3)-equivariant framework that builds the ECD spectrum from its underlying quantum-chemical quantities (Fig.~\ref{fig:arch}b).
PhysECD represents each excited state by its excitation energy and its electric and magnetic transition dipoles, and tracks how each quantity transforms under reflection.
The spectrum is then assembled from these quantities in closed form, and this architecture natively guarantees that enantiomers yield exactly sign-inverted spectra.
On the data side, we construct the Stereoisomer- and Conformer-resolved ECD
dataset (\textbf{SCECD}), the first ECD benchmark covering both multi-stereocentre molecules and conformer ensembles.
SCECD comprises 578,104 conformer spectra (1,156,208 after mirror
augmentation), obtained by CREST conformational search~\cite{crest}, DFT and
TDDFT calculations on 121,975 stereoisomers of 34,106 chiral parent molecules.
They carry one to five stereocentres, and are drawn from the COCONUT~\cite{coconut} and
LOTUS~\cite{lotus} natural-product databases.

Trained on these data, PhysECD predicts an ECD spectrum from a conformer geometry in milliseconds.
On CMCDS~\cite{cmcds}, a public ECD benchmark of single-stereocentre molecules, it outperforms existing predictors by a wide margin, with a per-molecule spectral Pearson correlation of 0.716 (mean) and 0.863 (median).
On SCECD, it reaches 0.753 (mean) and 0.912 (median) on Boltzmann-averaged ensemble spectra.
Among all candidate configurations of each parent molecule, it ranks the true configuration far above chance at every stereocentre count.
On 77 experimental spectra, including compounds with up to eleven stereocentres and twice as many atoms as the largest training molecules, it agrees with experiment as closely as TDDFT.
PhysECD can therefore serve as a fast, physics-constrained alternative to TDDFT calculations, offering a more efficient route to assigning the absolute configuration of complex chiral molecules.

\section{Results}\label{sec:results}

\subsection{A physics-constrained, parity-aware framework for ECD prediction}\label{sec:model}

PhysECD builds an ECD spectrum from the underlying quantum-chemical quantities that determine it (Fig.~\ref{fig:arch}b). 
Four modules act in sequence on a single
three-dimensional conformer. An \emph{E(3)-equivariant backbone} encodes the 
structure into per-atom features split by parity, with invariant ($0e$), polar ($1o$) and 
axial ($1e$) quantities occupying distinct and correctly transforming subspaces.
Multi-task \emph{atomic heads} map these features into the excitation energies 
and the electric and magnetic transition dipoles of the 20 lowest excited states. 
A parameter-free \emph{physics readout} pools them to the molecular level, and assembles 
the rotatory strengths in closed form. A \emph{Gaussian broadening module}
then converts the 20 discrete transitions into a continuous spectrum through
the standard Gaussian line shape, with quantum-chemically fixed constants (Methods).

Two properties are guaranteed by the architecture rather than learned from data.
First, PhysECD reproduces the defining chiroptical symmetry exactly.
Reflecting a molecule negates every predicted rotatory strength while leaving the excitation energies unchanged.
This follows from the parities of the two dipoles, a polar and an axial vector, whose scalar product reverses sign under reflection (Methods).
The prediction for an enantiomer is therefore the exact negative of the prediction for its partner, 
identically in the network parameters and at every stage of training. 
We also confirm this numerically in Supplementary Section 9. 
Second, as a direct consequence, mirror images add no new information to PhysECD.
Augmenting the training set with enantiomers leaves the training objective and its gradient unchanged, as we prove in Supplementary Section 10.

For the backbone, we evaluate the pretrained UMA foundation model~\cite{uma} and three E(3)-equivariant networks trained from scratch, Equiformer~\cite{equiformer}, DetaNet~\cite{detanet} and MACE~\cite{mace}.
Unlike the other three, the released UMA model is not equivariant to improper operations and cannot distinguish polar from axial vectors.
We modify it to bring it natively into the backbone contract without adding parameters (Supplementary Section 8).
Among the four, UMA performs best and is the default backbone.

Throughout this work, we report performance mainly as the per-molecule spectral Pearson correlation coefficient (PCC) between predicted and reference spectra over 180--400\,nm.
We summarize it by its mean, its median and the fraction of molecules with $\PCC>0$.
The sign of the PCC indicates whether the predicted handedness is correct.

\subsection{PhysECD outperforms existing ECD predictors on CMCDS}\label{sec:perf}

\begin{figure}[!t]
\vspace{\dimexpr\intextsep-1cm\relax}
\centering
\includegraphicsTG[\textwidth]{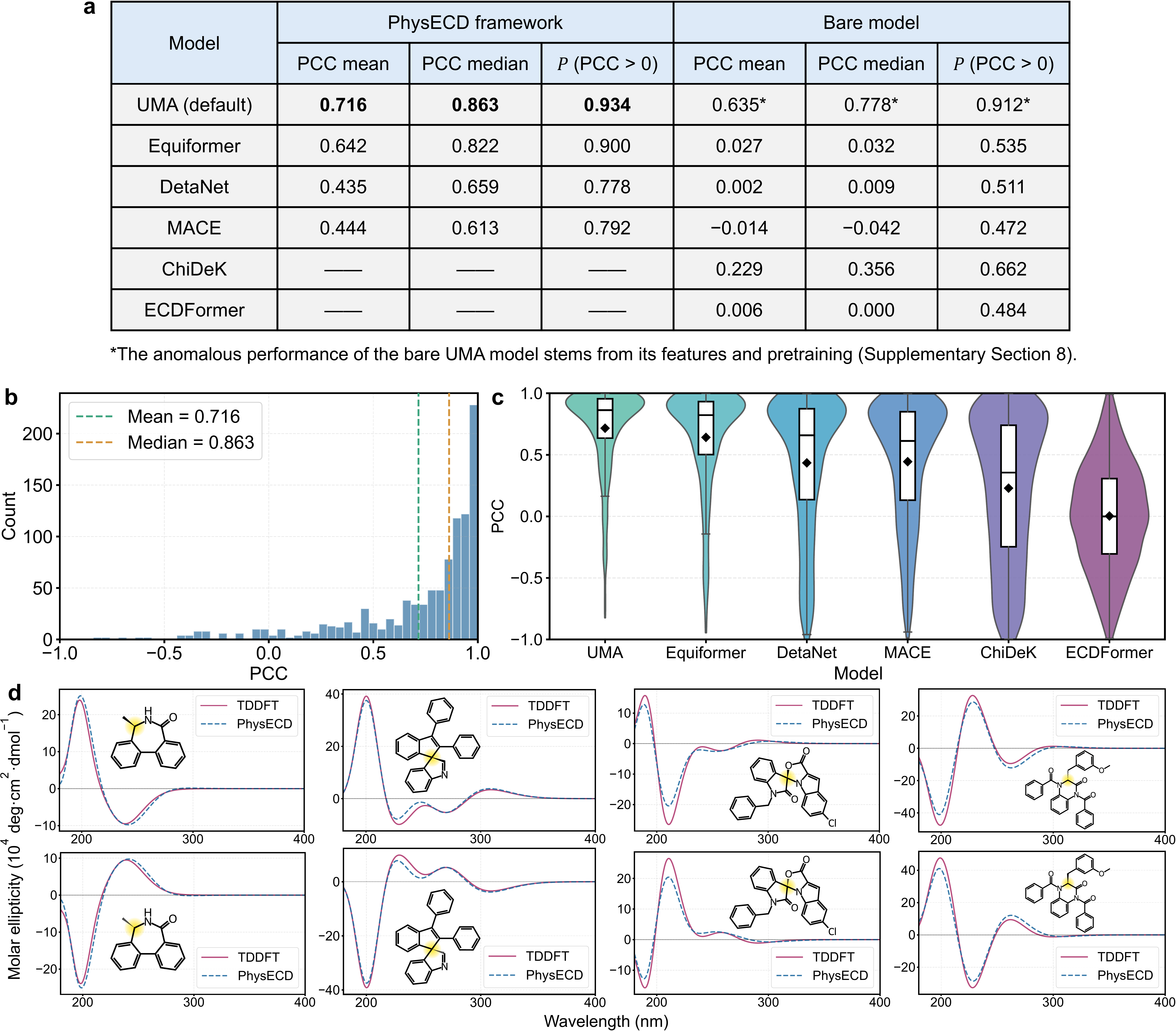}
\caption{\textbf{Performance of PhysECD on the CMCDS dataset.} \textbf{a},
Per-molecule spectral PCC over 180--400\,nm (higher is better), reported as its 
mean, its median and the fraction of test molecules with $\PCC>0$. ``PhysECD
framework'' couples each backbone with our physics-constrained architecture.
``Bare model'' denotes baselines trained to regress the spectrum directly. ChiDeK and ECDFormer cannot be coupled to the PhysECD readout and are therefore reported only as bare models (---).
\textbf{b}, Distribution of the per-molecule spectral PCC for PhysECD with the
UMA backbone. The mass concentrates near $+1$, with mean 0.716 (green dashed)
and median 0.863 (orange dashed). \textbf{c}, Per-molecule spectral PCC distributions
for the four PhysECD backbones, ChiDeK and ECDFormer. Boxes show the interquartile range
and median, diamonds the mean. \textbf{d}, Predicted (blue dashed) versus TDDFT
(red solid) ECD spectra for four pairs of representative enantiomers over
180--400\,nm. PhysECD tracks the sign, position and magnitude of the
dominant Cotton effects.}
\label{fig:cmcds}
\end{figure}

We first evaluate PhysECD on CMCDS~\cite{cmcds}, currently the only large-scale public benchmark of computed ECD spectra, which covers single-stereocentre organic molecules.
PhysECD predicts a spectrum in 9.1\,ms per conformer on a single GPU (Supplementary Section 6).
It attains a per-molecule spectral PCC of 0.716 (mean) and 0.863 (median), and assigns the correct absolute configuration for 93.4\% of test molecules (Fig.~\ref{fig:cmcds}a).
The distribution of PCC is strongly concentrated near $+1$ (Fig.~\ref{fig:cmcds}b).
Representative predictions track the sign, position and magnitude of the Cotton effects (Fig.~\ref{fig:cmcds}d).

The right-hand block of Fig.~\ref{fig:cmcds}a reports the same backbones stripped of the PhysECD framework.
These bare models are trained to regress the spectrum directly, with the same losses that PhysECD places on the spectrum.
ECDFormer and ChiDeK~\cite{chidek} are included only as bare-model baselines, because neither can be coupled to the PhysECD readout.
Without the framework, the bare Equiformer, DetaNet and MACE fall to chance, with mean PCCs near zero and a positive PCC for only about half of the test molecules.
These models fail because they give two enantiomers essentially the same spectrum.
Their invariant features do not change under reflection, and although 
the readout also receives the equivariant features, it does not learn to use them to distinguish enantiomers.
Replacing only the readout with one whose output reverses sign under reflection raises the mean PCC of the bare Equiformer from 0.027 to 0.579 (Supplementary Section 8).
The bare UMA is the exception.
Its features are not constrained by parity and can encode handedness.
Pretraining further improves its performance (Supplementary Section 8).
ChiDeK, whose input encodes the configuration of each stereocentre, likewise learns some handedness, reaching a mean PCC of 0.229.

We also evaluate PhysECD on ECDFormer's own data split and with its peak-based metrics. 
PhysECD still outperforms ECDFormer by a wide margin (Supplementary Section 5). 
Together, these results establish PhysECD's advantage over every prior predictor on CMCDS, but they also expose the limitations of this dataset.
CMCDS contains only single-stereocentre molecules and represents each molecule by a single geometry.
However, chiral molecules analysed in practice usually carry several stereocentres and exist as many conformers.
A model trained and validated on CMCDS therefore cannot yet support real experimental applications.
This gap motivated the construction of SCECD.

\subsection{SCECD extends ECD benchmarks to multiple stereocentres and conformer ensembles}\label{sec:dataset}

\begin{figure}[t]
\centering
\includegraphicsTG[\textwidth]{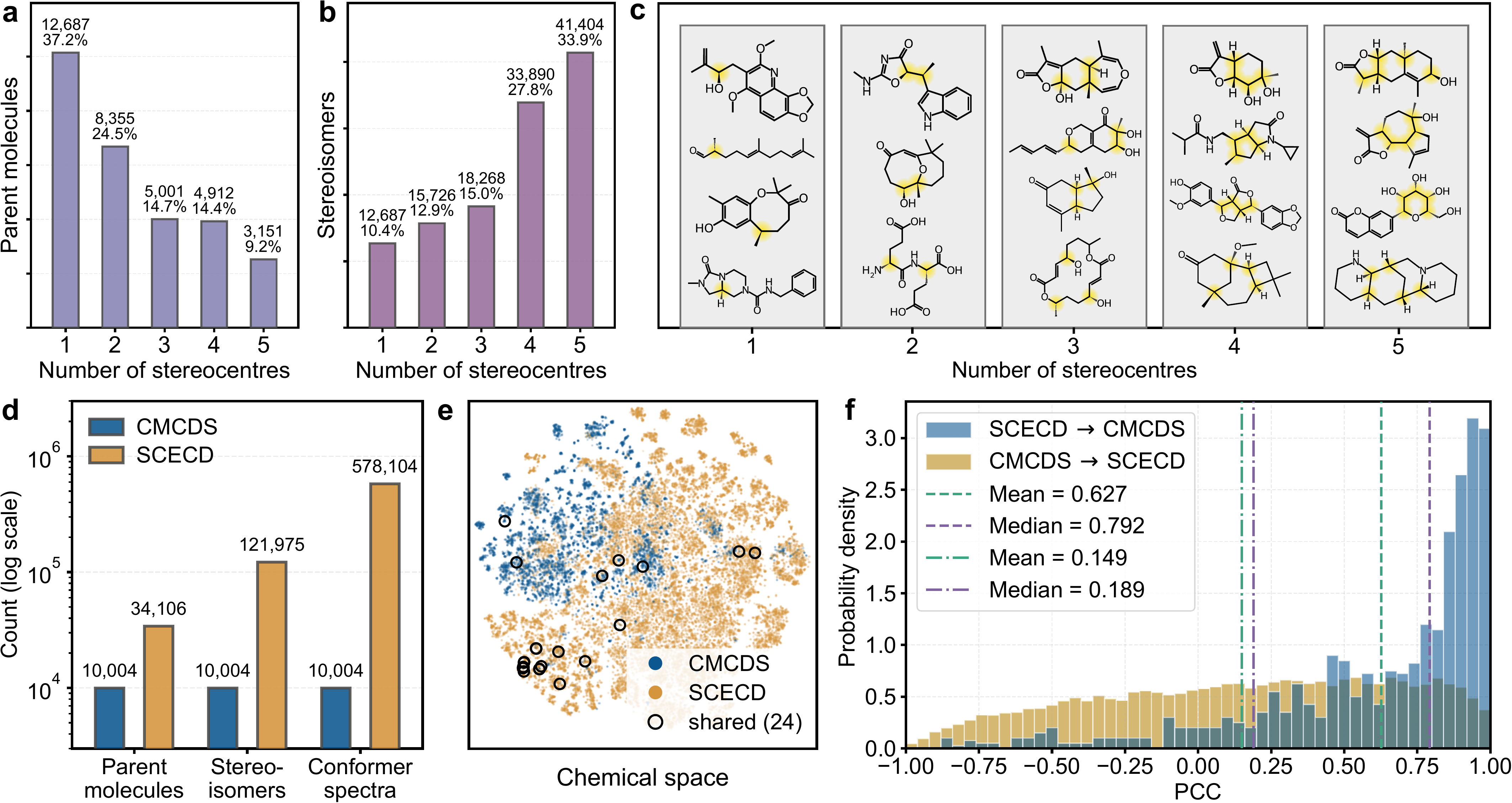}
\caption{\textbf{Overview of the SCECD dataset.} \textbf{a}, Number of parent molecules at each stereocentre count. SCECD contains 34,106 parent molecules carrying one to five stereocentres, and 62.8\% of them carry two or more. \textbf{b}, Number of stereoisomers at each stereocentre count. 
Because a parent with $n$ stereocentres has $2^{n-1}$ diastereomers, parents with four or five stereocentres make up only 23.6\% of all parents but contribute 61.7\% of the 121,975 stereoisomers. \textbf{c}, Representative parent molecules at each stereocentre count, with the
stereocentres highlighted. They illustrate the fused polycyclic terpenoid,
alkaloid, polyketide and glycoside skeletons that dominate SCECD. \textbf{d}, Size of SCECD relative to the CMCDS dataset at three levels (logarithmic axis). \textbf{e}, Chemical space comparison. The
two datasets occupy adjacent but largely disjoint regions. Black circles
mark the 24 structures that occur in both, 0.24\% of CMCDS and 0.07\% of SCECD (Supplementary Section 3). \textbf{f}, Cross-dataset
transfer. Per-molecule spectral PCC of PhysECD evaluated on the dataset it was not trained on.
SCECD\,$\rightarrow$\,CMCDS (blue) denotes an SCECD-trained model on the CMCDS test set, and CMCDS\,$\rightarrow$\,SCECD (gold) the reverse.
Vertical lines mark the mean (green) and median (purple) of each distribution, dashed for
SCECD\,$\rightarrow$\,CMCDS and dash-dotted for CMCDS\,$\rightarrow$\,SCECD.}
\label{fig:scecd}
\end{figure}

SCECD draws its molecules from the COCONUT~\cite{coconut} and LOTUS~\cite{lotus} natural-product databases.
It covers multi-stereocentre molecules (Fig.~\ref{fig:scecd}a), enumerates every diastereomer (Fig.~\ref{fig:scecd}b) and captures the Boltzmann conformer ensembles inherent to measured spectra.
Comprising 34,106 parent molecules, 121,975 stereoisomers and 578,104 conformers, SCECD contains nearly 58 times as many spectra as CMCDS (Fig.~\ref{fig:scecd}d).
To our knowledge, it is the first public ECD resource that resolves both diastereomers and conformers.
Construction details and dataset statistics are given in Supplementary Section 1 and Supplementary Section 2.

Both datasets were computed with the same quantum-chemical protocol (Methods). 
Therefore, a model trained on one dataset can be evaluated directly on the other.
The two datasets are nearly disjoint. Only 24
structures occur in both, and despite comparable molecular size, their Morgan
fingerprints~\cite{ecfp} occupy largely non-overlapping regions of chemical space.
As a result, each dataset is out of distribution with respect to the other (Fig.~\ref{fig:scecd}e; more details in Supplementary Section 3).

The cross-dataset evaluations are strongly asymmetric (Fig.~\ref{fig:scecd}f).
A model trained only on SCECD reaches a PCC of 0.627 (mean) and 0.792 (median) on the CMCDS test set, with a positive PCC for 91.2\% of molecules.
These values are close to the CMCDS-trained reference (Fig.~\ref{fig:cmcds}a).
By contrast, a model trained only on CMCDS collapses to 0.149, 0.189 and 61.4\% on the SCECD test set, far below the SCECD-trained reference.
This asymmetry shows that SCECD covers a distinct and more challenging region of chiral chemical space.
Training on SCECD is therefore sufficient to impart strong and robust out-of-distribution generalization to the models.

\subsection{PhysECD handles conformer ensembles and multiple stereocentres effectively on SCECD}\label{sec:scecd-perf}

\subsubsection{Conformer-resolved prediction and Boltzmann ensembles}\label{sec:scecd-ensemble}

\begin{figure}[t]
\centering
\includegraphicsTG[\textwidth]{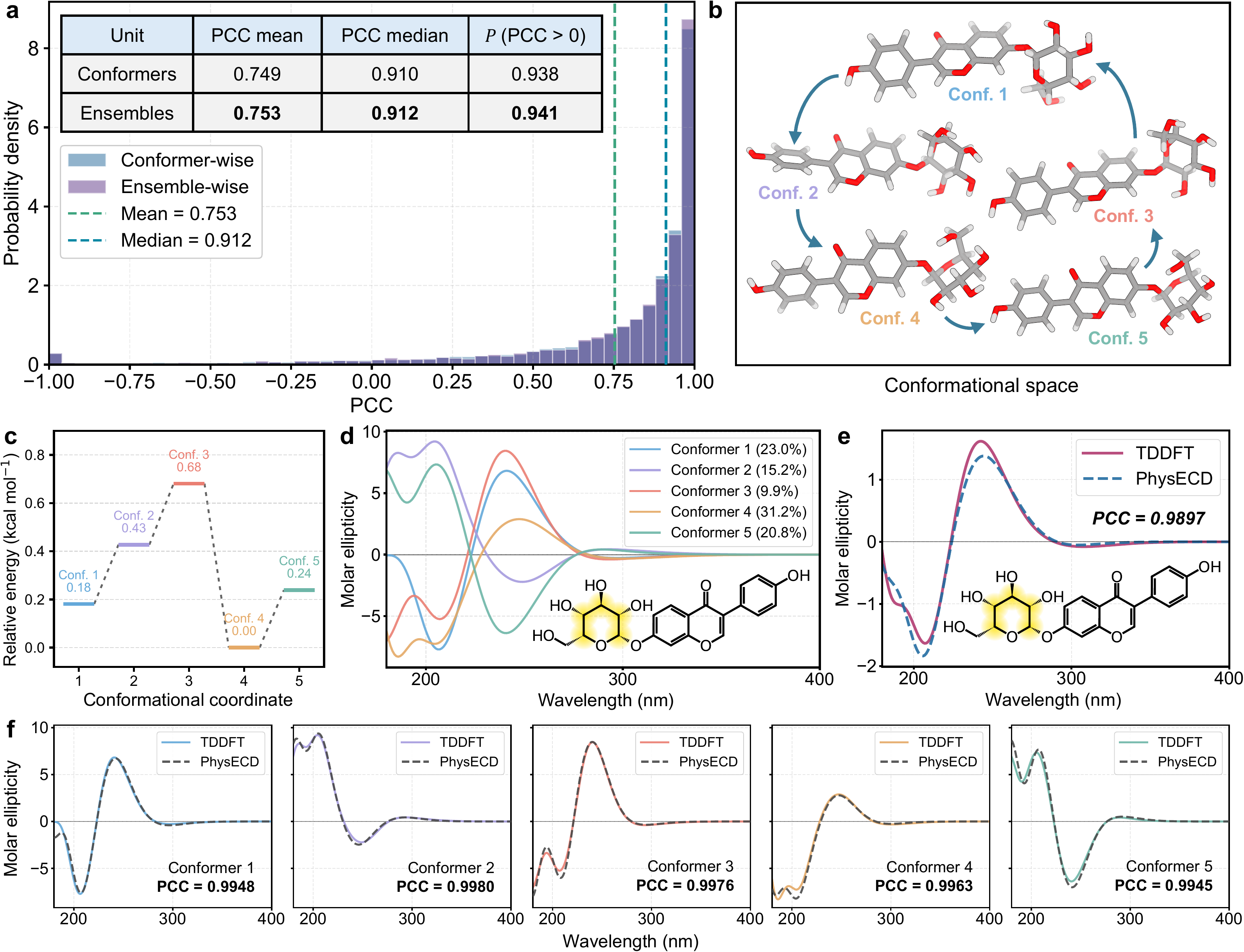}
\caption{\textbf{Conformer-resolved prediction and Boltzmann ensembles on
SCECD.} \textbf{a}, Distributions of PCC on the SCECD test set, for individual conformers and Boltzmann-averaged ensembles. The inset table gives the mean, median and $P(\PCC>0)$ of each. Dashed lines mark the ensemble mean and median.
\textbf{b}--\textbf{f}, One representative stereoisomer, with the five conformers coloured consistently throughout. \textbf{b}, The five optimized conformers.
\textbf{c}, Their relative DFT electronic energies. \textbf{d}, Their
TDDFT spectra overlaid, labelled with the Boltzmann populations at
298.15\,K. \textbf{e}, Predicted (blue dashed) and TDDFT (red solid)
Boltzmann-averaged ensemble spectra. \textbf{f}, Predicted and TDDFT
spectra of each individual conformer.}
\label{fig:ensemble}
\end{figure}

A measured ECD spectrum is not the spectrum of a single conformer.
It is a Boltzmann average over multiple conformers, which SCECD resolves explicitly.
Trained on SCECD with the UMA backbone, PhysECD reaches a PCC of 0.749 (mean) and 0.910 (median) on individual conformer spectra, with a positive PCC for 93.8\% of conformers.
Because PhysECD treats every conformer independently, the predicted conformer spectra can be Boltzmann-averaged in exactly the same way as TDDFT conformer spectra.
The resulting ensemble spectra reach a PCC of 0.753 (mean) and 0.912 (median), with a positive PCC for 94.1\% of stereoisomers (Fig.~\ref{fig:ensemble}a).
PhysECD is therefore accurate at the level where theory and experiment are actually compared, not only on individual conformers that are rarely observed in isolation.

A representative stereoisomer shows the Boltzmann-averaging process (Fig.~\ref{fig:ensemble}b--f).
Its five optimized conformers lie within 0.68\,kcal\,mol$^{-1}$ of one another and carry populations between 9.9\% and 31.2\%.
Therefore, none of them dominates the ensemble. 
Their TDDFT spectra differ in the sign and intensity of the two dominant Cotton effects, and PhysECD reproduces each spectrum with a PCC between 0.995 and 0.998.
After Boltzmann averaging, the predicted ensemble spectrum matches the TDDFT ensemble spectrum with a PCC of 0.990.
Further examples of ensemble predictions are given in Supplementary Section 14.

\subsubsection{Assigning absolute configuration across multiple stereocentres}\label{sec:mcaa}

The spectral PCC and absolute configuration assignment answer different questions.
The PCC measures how closely a predicted spectrum matches a reference, whereas absolute configuration assignment identifies which candidate configuration produced a measured spectrum.
With one stereocentre, or with several stereocentres of known relative configuration, the two questions nearly coincide. 
The two enantiomers have exactly opposite spectra, so the sign of the PCC identifies the true one.
With more than one stereocentre and an unknown relative configuration, the candidates also include diastereomers, whose spectra bear no fixed relation to one another.
Therefore, a positive PCC with one candidate no longer means that the correct configuration has been found.
To evaluate this task, we define the Top-$k$ Multi-stereocentre Configuration Assignment Accuracy (MCAA).
For every stereoisomer with $n$ stereocentres, we rank the predicted ensemble spectra of all $2^n$ candidate configurations by their PCC with a reference spectrum.
A query counts as a Top-$k$ hit when the true configuration falls within the top $k$ candidates (Methods).

\begin{figure}[t!]
\centering
\includegraphicsTG[\textwidth]{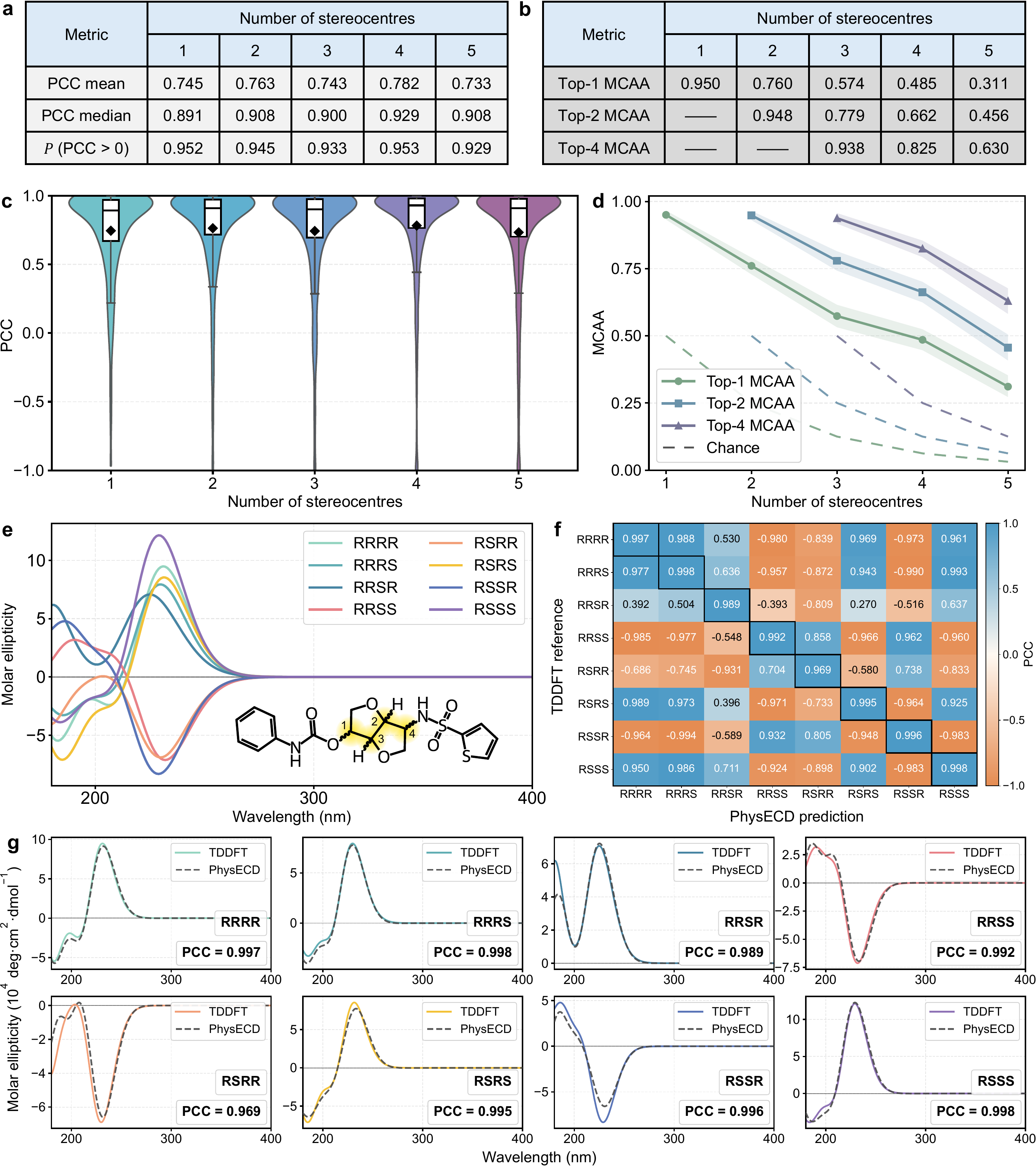}
\caption{\textbf{Performance and absolute-configuration assignment across
multiple stereocentres.} \textbf{a}, Ensemble spectral PCC metrics on the SCECD test set at
each stereocentre count. \textbf{b}, Top-1, Top-2 and Top-4 MCAA at
each stereocentre count. \textbf{c}, Ensemble PCC distributions at
each stereocentre count. Violins show the full distribution, boxes the
interquartile range and median, and diamonds the mean. \textbf{d}, The
line chart of \textbf{b}, with bootstrap 95\% confidence bands. Dashed lines of matching colour
give the $k/2^n$ chance rate. \textbf{e}--\textbf{g}, One four-stereocentre
parent molecule, with its eight diastereomers coloured consistently throughout.
\textbf{e}, Their TDDFT Boltzmann-ensemble spectra overlaid. The stereocentres are highlighted on the structure and numbered in the order of the configuration labels. Different configurations can yield highly similar spectra. \textbf{f}, PCC matrix
between the TDDFT spectrum and the PhysECD
prediction of each diastereomer. Black outlines mark the largest absolute PCC in each row. All queries are retrieved correctly. \textbf{g}, Predicted (dashed) and TDDFT (solid) ensemble spectra of each diastereomer.}
\label{fig:mcaa}
\end{figure}

On the SCECD test set, PhysECD reaches a Top-1 MCAA of 95.0\% for one-stereocentre and 76.0\% for two-stereocentre molecules, decreasing to 31.1\% at five (Fig.~\ref{fig:mcaa}b,d). 
This decline is expected, because the candidate set grows from 2 to 32 configurations.
However, the margin over chance widens, from 1.9 times the chance rate at one stereocentre to 9.9 times at five stereocentres. 
The Top-4 MCAA reaches 82.5\% for four-stereocentre and 
63.0\% for five-stereocentre molecules, 3.3 and 5.0 times the chance rate. Therefore, in most cases, the true configuration survives
a four- or eightfold reduction of the candidate set. Because every query is a TDDFT reference spectrum, these numbers quantify how
faithfully PhysECD reproduces the distinctions that TDDFT draws between
candidate configurations.

The lower Top-$k$ MCAA at higher stereocentre counts does not reflect less accurate spectral predictions.
The PCC remains stable from one to five stereocentres, with a mean between 0.733 and 0.782 and a median between 0.891 and 0.929 (Fig.~\ref{fig:mcaa}a,c).
A representative four-stereocentre molecule illustrates why a larger candidate set makes assignment harder (Fig.~\ref{fig:mcaa}e--g).
PhysECD reproduces the spectra of its eight diastereomers with PCCs between 0.969 and 0.998, and assigns the true configuration for every query.
Yet, the margin over the next-best candidate can be very small.
The TDDFT spectrum of RRRS has a PCC of 0.998 with the prediction for RRRS and 0.993 with that for RSSS, a margin of only 0.005.
The TDDFT spectrum of RSSR has a PCC of 0.996 with the prediction for RSSR and 0.994 with that for SSSR, a margin of only 0.002.
This example suggests that absolute configuration assignment at high stereocentre counts can be limited by the intrinsic similarity of candidate spectra.

\begin{figure}[t!]
\centering
\includegraphicsTG[\textwidth]{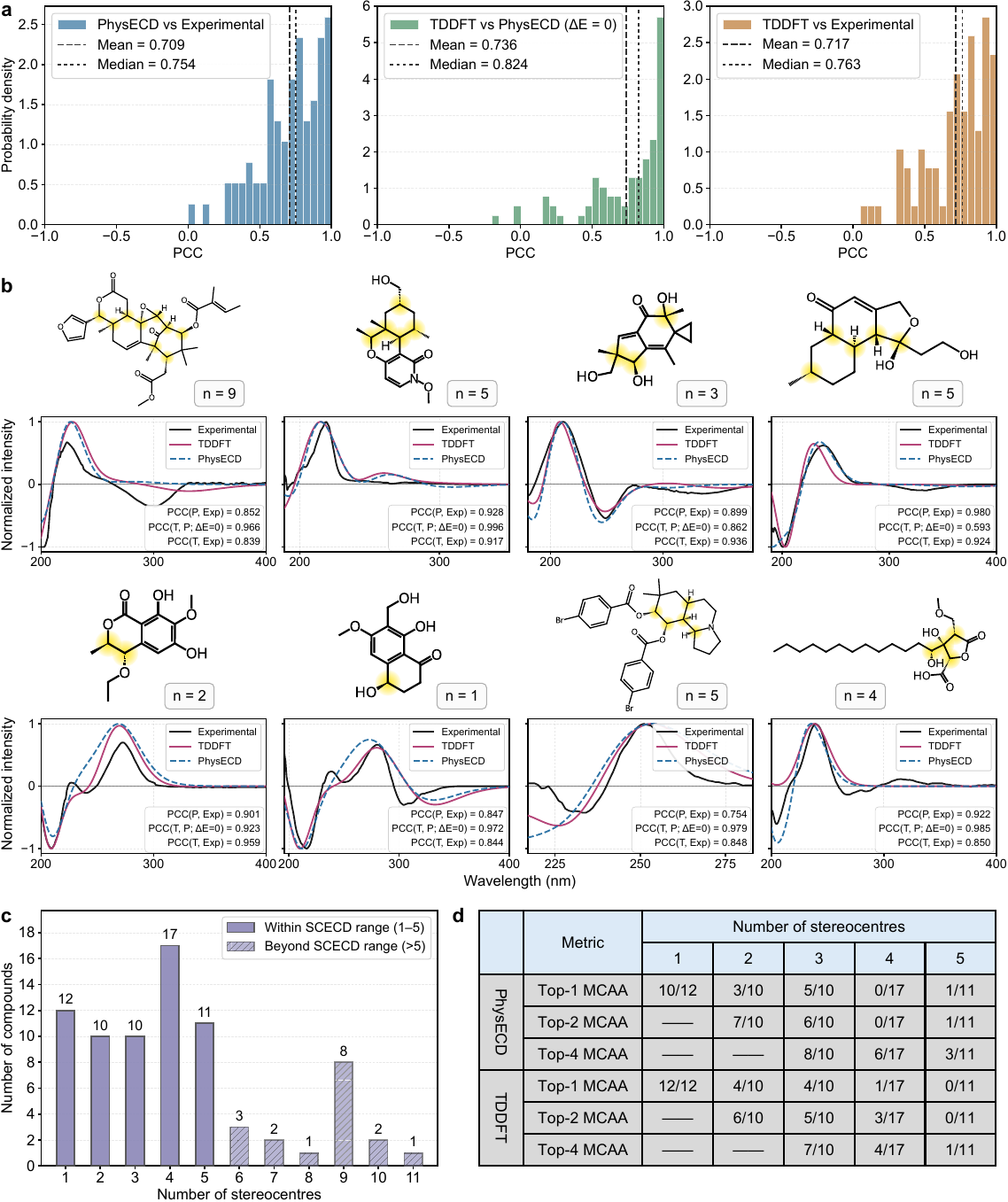}
\caption{\textbf{Validation on literature experimental ECD spectra.}
\textbf{a}, Distributions of spectral PCC over the 77 compounds. The panels show comparisons of PhysECD versus experiment, PhysECD versus TDDFT and TDDFT versus experiment. No energy shift is applied in the comparison between PhysECD and TDDFT. Dashed and dotted lines mark the mean and the median. 
\textbf{b}, Experimental (black), TDDFT (red) and PhysECD (blue dashed) spectra of eight representative compounds overlaid, each normalized to unit maximum absolute intensity. The stereocentres are highlighted on each structure. 
\textbf{c}, Number of compounds at each stereocentre count. Of these, 17 carry more than five stereocentres.
\textbf{d}, Top-$k$ MCAA evaluated against measured spectra for the 60 compounds with one to five stereocentres. Results are reported as successful assignments out of all compounds with that stereocentre count. Chance is $k/2^n$.}
\label{fig:exp}
\end{figure}

\subsection{Validation on experimental ECD spectra demonstrates the practical utility of PhysECD}\label{sec:experiment}

We next evaluated PhysECD against measured spectra. We collected 77 experimental
ECD curves from the natural-product literature and ensured none of them occurs in SCECD. These compounds are
more complex than the training molecules in both stereochemistry and size. They
carry one to eleven stereocentres, and 17 of them exceed the five-stereocentre
limit of SCECD (Fig.~\ref{fig:exp}c). They contain 28 to 105 atoms including
hydrogen, and 37 of them exceed the 50-atom limit of SCECD
(Supplementary Section 15). Before a calculated spectrum can be
compared with a measured one, its band positions need to be corrected. The
exchange--correlation functional systematically overestimates excitation
energies, and unmodelled solvent effects shift the bands further by
molecule-specific amounts. Therefore, each predicted or calculated ensemble was 
aligned using an excitation-energy shift fitted per parent molecule (Methods).

PhysECD reaches a spectral PCC against experiment of 0.709 (mean) and 0.754
(median), close to the 0.717 and 0.763 of TDDFT on the same spectra
(Fig.~\ref{fig:exp}a). On molecules larger and more stereochemically complex
than anything in the training set, PhysECD thus tracks experiment as closely as
the calculation it replaces. Compared directly without any energy shift,
PhysECD and TDDFT agree with a PCC of 0.736 (mean) and 0.824 (median). Spectral
overlays of eight representative compounds illustrate these correlations
(Fig.~\ref{fig:exp}b). PhysECD also matches TDDFT in configuration assignment. When the relative configuration is already known, the
two each identify the correct enantiomer for 59 of 77 compounds (76.6\%). When
it is unknown and all configurations of a skeleton are ranked against its
measured spectrum, they again perform comparably (Fig.~\ref{fig:exp}d).

Beyond their similar accuracy, PhysECD and TDDFT tend to fail on the same
compounds. Where both fail, their spectra still agree closely with each other
(Supplementary Section 15). Therefore, these shared failures may stem from 
the measured spectra or limitations of the TDDFT calculation method. 
Noise in the measured spectra is one such source, particularly in the four-stereocentre subset. 
There, strong noise below 230\,nm lowers the median PCC of the true configuration to
only 0.4, and both methods struggle (Supplementary Section 15).

\section{Discussion}\label{sec:discussion}

In this work, we developed PhysECD, a parity-aware E(3)-equivariant framework that builds an ECD spectrum from the excitation energy and the electric and magnetic transition dipoles of each excited state.
This framework ensures that enantiomers yield exactly sign-inverted spectra by construction.
Alongside the framework, we constructed SCECD, a conformer-resolved dataset of 578,104 spectra spanning the diastereomers of 34,106 natural products with up to five stereocentres.
PhysECD predicts spectra in milliseconds, with high fidelity for both individual conformers and Boltzmann ensembles.
For multi-stereocentre molecules in SCECD, it retrieves the true configuration from all candidate configurations at a success rate far above chance.
Furthermore, on 77 experimental spectra of more complex compounds, it agrees with experiment as closely as TDDFT.
PhysECD therefore offers a fast alternative to the TDDFT step in routine absolute configuration assignment, where the relative configuration is already known.
When the relative configuration is unknown, it can make exhaustive screening of candidate configurations more practical.

Despite these advances, several limitations remain and point to future work.
First, the reference data, and hence the model, describe isolated molecules in the gas phase.
Solvent effects are not considered, and extending SCECD with implicit- or explicit-solvent calculations~\cite{tomasi2005} could narrow the remaining gap with experiment.
Second, PhysECD replaces only the TDDFT step, while the conformational search and DFT geometry optimization remain.
Although replacing DFT with the highly efficient GFN2-xTB~\cite{gfn2xtb} lowers the spectral PCC only moderately (from $0.716$ to $0.675$, Supplementary Section 7), 
such semi-empirical energies are too coarse to set the Boltzmann weights.
More accurate machine learning interatomic potentials~\cite{aimnet2, mace-off} may eventually allow the DFT step to be bypassed.
Third, the data cover only central chirality. Axial and helical chirality~\cite{laplante2011, gingras2013} are not evaluated here, 
although the parity argument behind the framework does not depend on the origin of chirality and should carry over to these cases.
Fourth, the experimental spectra were digitized from many publications and recorded on different spectrometers. Their quality varies in ways that cannot be traced, limiting how precisely agreement with experiment can be measured.
Finally, PhysECD cannot effectively predict the per-state rotatory strengths and transition dipoles, which fail to generalize even under direct supervision (Supplementary Section 13).
Therefore, although the framework accelerates configuration assignment, it cannot replace TDDFT where the excited states themselves are the object of study. 
Addressing these limitations, together with broader experimental validation, would make PhysECD a more practical and powerful tool.

\section{Methods}\label{sec:methods}

\subsection{Problem formulation}\label{sec:methods-problem}
A single conformer with $N$ atoms is represented by its three-dimensional geometry $\mathcal{M} = \{(Z_A, \rr_A)\}_{A=1}^{N}$, where $Z_A$ and $\rr_A \in \mathbb{R}^3$ denote the atomic number and position of atom $A$, respectively.
Its ECD response is described by the $N_s=20$ lowest electronic excitations.
Each excited state $n$ has an excitation energy $E^n$, a velocity-form electric transition dipole $\muv^n$ and a magnetic transition dipole $\mv^n$.
These determine the velocity-form rotatory strength $R^n\propto\muv^n\!\cdot\!\mv^n/E^n$.
The observable spectrum is obtained by broadening the discrete set $\{(E^n,R^n)\}$ with a Gaussian line shape.
The learning task is the map $\mathcal{M}\mapsto\{(E^n,\muv^n,\mv^n)\}$, supervised through the excitation energies and the broadened spectrum.
The map from these quantities to the spectrum is fixed by physics and contains no free parameters.
All spectra in this work are evaluated on a 180--400\,nm grid at 1\,nm spacing (221 points), which covers the window accessible to standard ECD instrumentation.

\subsection{Parity of the quantities that determine the ECD spectrum}\label{sec:methods-parity}
ECD measures the difference $\Delta\varepsilon$ between the molar absorption coefficients of left- and right-circularly polarized light.
Within the electric-dipole--magnetic-dipole approximation, the intensity of band $n$ is governed by its rotatory strength $R^n$~\cite{rosenfeld1929, pescitelli2016}.
The rotatory strength depends on the electric and magnetic transition dipoles $\muv^n$ and $\mv^n$ only through their scalar product $\muv^n\cdot\mv^n$.
Here $\muv^n$ is the velocity-form electric transition dipole, built from the electronic linear momentum $\hat{\muv}\propto\sum_i\mathbf{p}_i$, where $i$ runs over the electrons.
The magnetic transition dipole $\mv^n$ is built from the electronic angular momentum $\hat{\mv}\propto\sum_i\rr_i\times\mathbf{p}_i$.

Under an orthogonal transformation $Q\in O(3)$, the position $\rr_i$ and momentum $\mathbf{p}_i$ of each electron transform as polar vectors, $\mathbf{v}\mapsto Q\mathbf{v}$.
The velocity-form electric transition dipole is therefore also a polar vector, $\muv^n\mapsto Q\muv^n$.
A cross product of two polar vectors acquires an additional factor $\det Q$, so the magnetic transition dipole is an axial vector, $\mv^n\mapsto\det(Q)\,Q\,\mv^n$.
The two dipoles thus transform identically under proper rotations ($\det Q=+1$) and differ only under improper operations ($\det Q=-1$), such as reflections.

An enantiomer is generated by an improper operation, for example the reflection $P=\mathrm{diag}(1,1,-1)$ through the $xy$ plane, for which $\det P=-1$.
Applying $P$ to a conformer maps $\rr_A\mapsto P\rr_A$.
The excitation energies are unchanged, and the two dipoles follow the transformation laws above,
\begin{equation}\label{eq:parity-map}
E^n\mapsto E^n,\qquad
\muv^n\mapsto P\,\muv^n,\qquad
\mv^n\mapsto \det(P)\,P\,\mv^n .
\end{equation}
Because $P$ is orthogonal, $(P\muv^n)\cdot(\det(P)\,P\mv^n)=\det(P)\,\muv^n\cdot\mv^n$, and the rotatory strength therefore transforms as
\begin{equation}\label{eq:pseudoscalar}
R^n\mapsto\det(P)\,R^n=-R^n .
\end{equation}
It is thus a pseudoscalar, invariant under rotations and translations but sign-reversing under reflections.
Broadening transitions with unchanged energies and negated rotatory strengths yields the exactly inverted spectrum, $\Delta\varepsilon\mapsto-\Delta\varepsilon$. This is why enantiomers have mirror-image ECD spectra.
Therefore, Eq.~\eqref{eq:pseudoscalar} is the physical basis for assigning absolute configuration by ECD.

\subsection{Construction of the SCECD dataset}\label{sec:methods-scecd}
\paragraph{Molecule selection and diastereomer enumeration.}
Candidate structures were drawn from the COCONUT~\cite{coconut} and LOTUS~\cite{lotus} open natural-product collections.
Eligible molecules carried one to five tetrahedral stereocentres and at most 50 atoms including hydrogens.
To keep abundant skeletons from crowding out rare ones, the eligible pool was thinned to at most two molecules per Bemis--Murcko scaffold~\cite{bemis1996}.
The full selection pipeline and its intermediate counts are given in~Supplementary Section 1.
For each retained parent molecule with $n$ stereocentres, all $2^{n-1}$ diastereomers were enumerated.
This produced 34,106 parent molecules and 121,975 stereoisomers.

\paragraph{Conformational search.}
Each stereoisomer underwent a CREST conformer--rotamer ensemble search~\cite{crest} at the GFN2-xTB level~\cite{gfn2xtb} in the gas phase, using the \texttt{mquick} metadynamics mode with an 8\,kcal\,mol$^{-1}$ energy window.
Because xTB relaxation can invert a labile stereocentre, the conformers were checked for stereochemical integrity in order of increasing energy.
The lowest-energy conformers that passed this check were retained, up to five conformers per stereoisomer.

\paragraph{Electronic-structure calculations.}
All DFT and TDDFT calculations were performed with Gaussian~\cite{gaussian09}.
Every retained conformer was optimized at the B3LYP/6-31G(d) level~\cite{becke1993, lyp1988, stephens1994} in the gas phase.
The 20 lowest excited states were then computed by TDDFT at the CAM-B3LYP/6-31G(d) level~\cite{camb3lyp, furche2002}.
These settings match the calculation protocol of the CMCDS dataset~\cite{cmcds}.
The formal charge was derived for each molecule, and every system was treated as a closed-shell singlet.
For each conformer, we stored the optimized geometry, the 20 excitation energies, the electric and magnetic transition dipoles and the rotatory strengths.
Gaussian reports the electric transition dipole and the rotatory strength in both the length and the velocity gauge.
We retained the origin-independent velocity form for both the dataset and the framework.
Calculations that failed to converge or did not terminate normally were discarded, leaving 578,104 conformer spectra.

\paragraph{Mirror augmentation.}
The enantiomer of every conformer was generated analytically by reflecting the
$z$ coordinates and transforming the quantities by Eq.~\eqref{eq:parity-map} ($\mu^n_z\mapsto-\mu^n_z$; $m^n_x,m^n_y\mapsto-m^n_x,-m^n_y$; $R^n\mapsto-R^n$; $E^n$ unchanged). 
This process doubled the dataset to 1,156,208 conformer spectra without further calculation. 
Mirror partners were always assigned to the same data split as their parents.

\paragraph{Conformer populations and ensemble spectra.}
Conformer spectra are combined into ensemble spectra using Boltzmann weights at $T=298.15$\,K, derived from the CAM-B3LYP electronic energies at the B3LYP-optimized geometries.
Because distinct CREST starting structures frequently relax into the same DFT minimum, near-duplicate optimized conformers are first merged.
Two conformers are treated as duplicates when their heavy-atom root-mean-square deviation (RMSD) after Kabsch alignment~\cite{kabsch1976} is at most 0.01\,\AA{}, and only the lowest-energy member of each duplicate group is kept.
The weight of conformer $c$ is then
\begin{equation}
w_c=\frac{\exp(-\Delta \mathcal{E}_c/k_BT)}{\sum_{c'}\exp(-\Delta \mathcal{E}_{c'}/k_BT)},
\end{equation}
where $\Delta\mathcal{E}_c$ is the electronic energy of conformer $c$ relative to the lowest-energy conformer and $k_B$ is the Boltzmann constant.
The same weights are applied to the TDDFT and predicted conformer spectra.

\subsection{E(3)-equivariance and parity-aware equivariant neural networks}\label{sec:methods-equiv}
Let a group $G$ act on an input space $\mathcal{X}$ through transformations
$\rho_g:\mathcal{X}\to\mathcal{X}$ and on an output space $\mathcal{Y}$ through
$\rho'_g:\mathcal{Y}\to\mathcal{Y}$. A function $\phi:\mathcal{X}\to\mathcal{Y}$ is \emph{equivariant} with respect to $G$ if
\begin{equation}\label{eq:equivariance}
\phi\big(\rho_g(x)\big)=\rho'_g\big(\phi(x)\big),\qquad g\in G,\ x\in\mathcal{X},
\end{equation}
and \emph{invariant} if the output is unchanged,
$\phi(\rho_g(x))=\phi(x)$. Equivariance does not require a prediction to stay unchanged when the input is transformed, 
only to transform as the corresponding physical quantity does. 
For a molecular property predicted from three-dimensional coordinates, the relevant group is the Euclidean group E(3). It is generated by translations and the orthogonal group $O(3)$ of rotations and reflections. 
Translation equivariance follows from building every geometric feature from relative displacements $\rr_{AB}=\rr_B-\rr_A$. 
Therefore, the remaining requirement is $O(3)$-equivariance, and here a model must decide whether to account for parity.

The irreducible representations (irreps) of $O(3)$ are labelled by a degree $\ell\in\{0,1,2,\dots\}$ and a parity $p\in\{e,o\}$.
A feature of degree $\ell$ has $2\ell+1$ components, and a rotation transforms these components.
Under spatial inversion, the feature is multiplied by $+1$ if $p=e$ (even), and by $-1$ if $p=o$ (odd).
Equivariant networks propagate such features and combine them through tensor products.
Every feature produced by these tensor products also carries an irrep label~\cite{tfn,e3nn}.
Architectures built in this way are now widely used for molecular property prediction~\cite{nequip,equiformer,mace}.
Because the irrep labels are attached to the features themselves, they can classify different types of physical quantities directly.
An invariant scalar is $0e$ and a pseudoscalar is $0o$.
A polar vector, such as a position or an electric dipole, is $1o$, whereas an axial vector, such as an angular momentum or a magnetic dipole, is $1e$.

The parity label separates two otherwise similar families of models.
An SE(3)-equivariant network~\cite{se3transformer, painn} does not distinguish the two parities.
By contrast, a parity-aware, fully E(3)-equivariant network keeps $o$ and $e$ features separate throughout.
An improper operation acts on the two kinds of features with opposite signs, and quantities of either parity can be emitted as separate outputs. 
A third and more restrictive family comprises invariant architectures, such as
SchNet~\cite{schnet}. They discard directional information and cannot represent vectors. 
Among the three families, only the parity-aware E(3)-equivariant model can natively reproduce the sign inversion of ECD under reflection. 
This is the property PhysECD requires of its backbone.

\subsection{The PhysECD framework}\label{sec:methods-framework}
\paragraph{Backbone contract.} An E(3)-equivariant backbone maps $(Z_A,\rr_A)$ to
per-atom features split into invariant scalars $S_A\in\mathbb{R}^{d}$ ($0e$) and
equivariant tensors $T_A$ that carry both parities (at least $1o$ and $1e$). 
Every backbone follows a fixed contract, emitting $(S_A,T_A)$ and exposing the irreps of $T_A$. 
Thus, different backbones are interchangeable behind an identical downstream head.

\paragraph{Multi-task atomic heads.}
Three heads act on each atom and produce per-state quantities. 
A scalar multilayer-perceptron head maps $S_A$ to per-atom excitation-energy contributions $E^n_A$ for the $N_s=20$ states. 
An equivariant $1o$ head maps $T_A$ to $N_s$ electric transition dipoles $\muv^n_A$, and an equivariant $1e$ head maps
$T_A$ to $N_s$ magnetic transition dipoles $\mv^n_A$.

\paragraph{Parameter-free physics readout.} A single readout aggregates the
atomic predictions into molecular quantities, with the pooling dictated by
physics (energies are intensive, dipoles extensive):
\begin{gather}
\muv^n=\sum_A\muv^n_A,\qquad
\mv^n=\sum_A\mv^n_A,\qquad
E^n=\mathrm{softplus}\bigg(\frac{1}{N}\sum_A E^n_A\bigg)+3.0\ \mathrm{eV},\\
R^n=C_R\,\frac{\muv^n\cdot\mv^n}{E^n}\quad(10^{-40}\,\mathrm{cgs}).\label{eq:R}
\end{gather}
The softplus function, $\mathrm{softplus}(x)=\ln(1+e^{x})$, is always positive.
Together with the 3.0\,eV offset, it keeps $E^n$ above 3.0\,eV.
This floor matches the lowest excitation energy in the reference data and the 400\,nm edge of the evaluation window, and it keeps the division in Eq.~\eqref{eq:R} well defined. 
The constant $C_R=6414.135151$ is the unit conversion of the velocity-form rotatory strength (see Supplementary Section 11 for detailed derivations). 
The readout has no learnable parameters. 
An alternative that builds $R^n$ from one predicted pseudoscalar per state, instead of two dipoles, is compared with it in~Supplementary Section 12.

\paragraph{Gaussian broadening.} 
The discrete transitions are broadened into a continuous spectrum with a Gaussian line shape of width $\sigma=0.4$\,eV. The broadening is performed on the energy grid $\tilde e=1240/\lambda$ ($\lambda$ in nm), and the resulting $\Delta\varepsilon$ is then converted to molar ellipticity $[\theta]$:
\begin{equation}
\Delta\varepsilon(\tilde e)=\frac{1}{2.296\times10^{1}\,\sigma\sqrt{\pi}}\sum_{n=1}^{N_s}E^n R^n\exp\!\Big[\!-\big((\tilde e-E^n)/\sigma\big)^2\Big],\qquad [\theta]=3298.2\,\Delta\varepsilon .
\end{equation}
Here $\sigma$ is the $1/e$ half-width in the Stephens--Harada convention~\cite{stephensharada2010}, corresponding to a full width at half maximum (FWHM) of $2\sqrt{\ln 2}\,\sigma\approx0.67$\,eV.
This is the default broadening of the widely used Multiwfn package~\cite{multiwfn}, and $\sigma$ lies within the 0.2--0.4\,eV range commonly adopted for TDDFT ECD simulation. 
The same broadening is applied to the predicted and TDDFT spectra across all experiments. 
The constants $2.296\times10^{1}$ and $3298.2$ are fixed by standard quantum-chemistry conventions and are derived in~Supplementary Section 11.
The whole map is differentiable in $(E^n,R^n)$, and thus gradients from a spectrum-level loss propagate back to the per-state quantities and the network.

\paragraph{Training objective.} The loss is a weighted sum of four complementary terms:
\begin{gather}
\mathcal{L} = w_{\text{shape}}\,\mathcal{L}_{\text{shape}}
+ w_{\text{mag}}\,\mathcal{L}_{\text{mag}}
+ w_{E}\,\mathcal{L}_{E}
+ w_{\text{reg}}\,\mathcal{L}_{\text{reg}},\\[2pt]
\mathcal{L}_{\text{shape}} = 1-\PCC\big([\theta]_{\text{pred}},[\theta]_{\text{TDDFT}}\big),\quad
\mathcal{L}_{\text{mag}} = \mathrm{SmoothL1}\big(10^{-4}[\theta]_{\text{pred}},\,
10^{-4}[\theta]_{\text{TDDFT}}\big),\quad \\
\mathcal{L}_{E} = \mathrm{MSE}\big(E^n_{\text{pred}},E^n_{\text{TDDFT}}\big),\quad
\mathcal{L}_{\text{reg}} = \frac{1}{3N_s}\sum_{n}\big(\|\muv^n\|^2+\|\mv^n\|^2\big),
\end{gather}
with $w_{\text{shape}}=w_{\text{mag}}=1.0$, $w_{E}=0.1$ and
$w_{\text{reg}}=0.2$ throughout.
MSE denotes the mean squared error, and SmoothL1 denotes the Huber loss with unit threshold.

The four terms constrain both the macroscopic spectrum and the per-state quantities. 
The shape term $\mathcal{L}_{\text{shape}}$ fits the spectral envelope, including the signs and relative line shapes of the Cotton effects, and is insensitive to the overall amplitude.
The magnitude term $\mathcal{L}_{\text{mag}}$ sets the absolute intensity scale, and the energy term $\mathcal{L}_{E}$ anchors the 20 excitation energies.
The minimum-norm regularizer $\mathcal{L}_{\text{reg}}$ constrains the dipole degrees of freedom that the spectrum leaves undetermined, which helps the shape term converge further.
A loss ablation is reported in~Supplementary Section 12.

Neither the transition dipoles nor the rotatory strengths are supervised directly, because they do not generalize.
Supervising them lowers the training loss, but the validation loss rises and the spectral accuracy falls (Supplementary Section 13).

\paragraph{Exact parity equivariance.} Every stage above preserves the irrep labels, because the dipole heads are equivariant linear maps, the pooling is linear and the readout is a fixed closed-form expression.
A reflection $P$ therefore transforms $\muv^n$ and $\mv^n$ as in Eq.~\eqref{eq:parity-map}.
As a result, Eq.~\eqref{eq:pseudoscalar} is guaranteed by the architecture, and the whole ECD prediction network $f_\theta$ satisfies
\begin{equation}\label{eq:parity-equivariance}
f_\theta(P\cdot\mathcal{M})=-f_\theta(\mathcal{M})
\end{equation}
for every value of the parameters $\theta$, at initialization and throughout training.

\subsection{Evaluation metrics}\label{sec:methods-metrics}
\paragraph{Spectral correlation.} The primary metric is the per-molecule spectral Pearson correlation coefficient between the predicted and reference spectra on the 180--400\,nm grid.
We summarize it by its mean, its median and the fraction of molecules with $\PCC>0$.
The PCC is insensitive to a global scale factor and to a constant offset, and its sign reports whether the predicted handedness is correct.

\paragraph{Top-$k$ multi-stereocentre configuration assignment accuracy.} For a parent molecule $g$ with $n$ stereocentres, we consider all of its $2^n$ absolute configurations as candidates.
PhysECD predicts the spectrum of every conformer of each candidate, and these spectra are Boltzmann-averaged into the predicted ensemble spectrum $\hat{\mathbf{y}}_{g,j}$ of candidate $j$.
Each stereoisomer $i$ of $g$ serves as a query, with its reference ensemble spectrum $\mathbf{y}_{g,i}$.
The candidates are ranked by $s_{g,ij}=\PCC(\mathbf{y}_{g,i},\hat{\mathbf{y}}_{g,j})$, and the rank of the true configuration is
\begin{equation}
r_{g,i}=\sum_{j=1}^{2^n}\mathbb{I}\big(s_{g,ij}\ge s_{g,ii}\big).
\end{equation}
A query is a Top-$k$ hit when $r_{g,i}\le k$, and the reported accuracy is the fraction of hits among the $Q_n$ queries with $n$ stereocentres,
\begin{equation}
\text{Top-}k\ \text{MCAA}_n=\frac{1}{Q_n}\sum_{g,i}\mathbb{I}\big(r_{g,i}\le k\big).
\end{equation}
Confidence intervals are obtained by resampling parent molecules with replacement 5,000 times and taking the 2.5\% and 97.5\% quantiles~\cite{efron1979}.
The chance rate for a random shortlist of size $k$ is $k/2^n$.

\subsection{Training setup and hyperparameters}\label{sec:methods-setup}
\paragraph{Dataset preprocessing and split protocol.} We first deduplicate CMCDS to 10,004 distinct parent molecules (Supplementary Section 4).
Both datasets are then split 90/5/5 at the level of the parent molecule, with every enantiomer, stereoisomer and conformer of a parent confined to one subset.
This gives 9,003/500/501 parents for CMCDS and 30,695/1,705/1,706 for SCECD, the latter yielding 517,764/29,811/30,529 conformer spectra.
PhysECD is trained and evaluated on these non-augmented splits, because mirror augmentation is redundant for a parity-aware model (Supplementary Section 10).
By contrast, the baselines require mirror augmentation to learn chirality. Therefore, they are evaluated on the within-split enantiomer expansion, 18,006/1,000/1,002 molecules for CMCDS.
Our split protocol differs from the split released with ECDFormer, which shuffles a flat list of mirror-augmented spectra and slices it directly without grouping by parent molecule (Supplementary Section 4).
Results on that split are reported in~Supplementary Section 5.

\paragraph{Hyperparameters.} The three E(3)-equivariant backbones trained from scratch share the same hyperparameters.
We use AdamW~\cite{adam, adamw} with a weight decay of $10^{-5}$ and a learning rate $\eta$ scaled with the global batch size $B$ as $\eta=\eta_0\sqrt{B/B_0}$, where $\eta_0=10^{-3}$ and $B_0=64$.
At $B=512$, used throughout, this gives $\eta\approx2.8\times10^{-3}$. 
The schedule is a 5-epoch linear warmup followed by cyclic cosine annealing~\cite{sgdr} with a period of 398 epochs.
Training runs for 1,000 epochs with gradient-norm clipping at 1.0 and random seed 42, and the model is selected by the validation shape loss.
The UMA backbone is instead fine-tuned from the released \texttt{uma-s-1p1} checkpoint.
Fine-tuning uses AdamW with a weight decay of $10^{-2}$, a constant learning rate of $2\times10^{-4}$ for the backbone and heads, no warmup, gradient-norm clipping at 10.0, random seed 42 and a global batch size of 64.
It runs for 100 epochs on both CMCDS and SCECD.
Hyperparameters of the other models and the full configurations are listed in~Supplementary Section 8.

\subsection{Validation against experimental spectra}\label{sec:methods-experiment}
\paragraph{Spectrum digitization.} The 77 experimental spectra were digitized from the figures of the source articles (Supplementary Section 15).
Both axes were calibrated by an affine fit to the visible tick marks, with the wavelength range resolved at the plot borders.
The experimental trace was separated from axes, legends and any calculated curve by colour and line-style masking with explicit exclusion regions.
It was then followed pixel column by pixel column and resampled onto a uniform 1\,nm grid.
Every curve had to pass automatic checks on grid spacing, sample count and ordinate range, as well as a visual check that overlaid the reconstruction on the source image.

\paragraph{Reference calculations.} The TDDFT ensembles of the literature
compounds were computed as for SCECD, except that up to 20 conformers were
retained per stereoisomer.

\paragraph{Excitation-energy alignment.} Each predicted or calculated ensemble spectrum was aligned to the measured one by a rigid shift $\Delta E\in[-1,1]$\,eV added to all 20 excitation energies of every conformer.
The shift leaves the rotatory strengths and Boltzmann weights unchanged.
It was fitted per parent molecule and separately for PhysECD and TDDFT. The direct comparison between PhysECD and TDDFT used no shift.
For the spectral comparison, $\Delta E$ was chosen to maximize the $\PCC$ with the measured spectrum.
For configuration assignment, all candidates of a compound shared one shift, because enantiomers have identical excitation energies and diastereomers differ only marginally.
This shift was estimated without prior knowledge of the true configuration,
\begin{equation}\label{eq:consensus-shift}
\Delta E^{*}=\operatorname*{arg\,max}_{\Delta E}\
\operatorname{median}_{j}\big|\PCC\big(\mathbf{y}_{\mathrm{exp}},\,\hat{\mathbf{y}}_j(\Delta E)\big)\big| ,
\end{equation}
where $\hat{\mathbf{y}}_j(\Delta E)$ is the shifted ensemble spectrum of candidate $j$ and $\mathbf{y}_{\mathrm{exp}}$ is the measured spectrum.
The median prevents any single candidate from setting the shift.
At this shift, the enantiomer assignment accuracy is the fraction of the 77 compounds whose true configuration gives $\PCC>0$.
Top-$k$ MCAA was computed from the same spectra for the 60 compounds with one to five stereocentres.

\backmatter

\section*{Data availability}
The CMCDS dataset is available via Figshare at \url{https://doi.org/10.6084/m9.figshare.28805909}. The SCECD dataset introduced in this study, its raw Gaussian output and the curated literature experimental ECD validation set are deposited on Zenodo at \url{https://doi.org/10.5281/zenodo.22845673}, \url{https://doi.org/10.5281/zenodo.22846559} and \url{https://doi.org/10.5281/zenodo.22847136}, respectively. They are accessible to editors and reviewers during peer review and will be made publicly available upon publication.

\section*{Code availability}
The PhysECD code is archived on Zenodo at \url{https://doi.org/10.5281/zenodo.22951956} and is accessible to editors and reviewers during peer review. It will be made publicly available on GitHub and Zenodo upon publication.

\renewcommand{\bibfont}{\normalfont\normalsize}%
\setlength{\bibsep}{0pt}%
\renewcommand{\bibcleanup}{\vskip-\lastskip\vskip0pt}%
\bibliography{sn-bibliography}

\section*{Acknowledgements}
This work was supported by the National Natural Science Foundation of China (Nos. 62272300, 62672312, 92570205 and 92461313) and the Chuangqi Program of Shanghai Innovation Institute, Shanghai Municipal Education Commission. We acknowledge support from the Open Research Fund of Suzhou Laboratory (No. SZLAB-1508-2024-TS017). This work was also supported by the Basic Science Research Program (Priority Research Institute) through the National Research Foundation (NRF) of Korea funded by the Ministry of Education (No. RS-2021-NR060138), by the Creative Materials Discovery Plus Program through the NRF of Korea funded by the Ministry of Science and ICT (No. RS-2026-25596523) and by the National Research Lab 2.0 through the NRF grant funded by the Korea government (MSIT and MOE) (No. RS-2025-16063688). We thank Shanghai Innovation Institute and School of Computer Science, Shanghai Jiao Tong University, for providing GPU and CPU resources.

\section*{Author contributions}
Y.J., L.C., T.Z. and Y.Y. conceived and designed the study. Y.J. and L.C. constructed the SCECD dataset. Y.J. developed and implemented the PhysECD framework, performed the experiments and analysed the results. L.C., R.S. and L.B. contributed to the performance optimization and codebase of the framework. Y.J. and L.C. curated the experimental ECD spectra from the literature. Y.J. drafted the manuscript and prepared the figures. L.C., R.S., Z.L., D.H.K., T.Z. and Y.Y. revised the manuscript and figures. T.Z. and Y.Y. supervised the project and provided the computational resources.

\section*{Competing interests}
The authors declare no competing interests.

\end{document}